\documentstyle[12pt,epsf]{article}

\begin{document}

\renewcommand{\thefootnote}{\fnsymbol{footnote}}

\pagestyle{empty}
\pagenumbering{arabic}

\begin{flushright}
ZU-TH 22/95\\
TTP95-31\\
\today\\[15mm]
\end{flushright}

\begin{center}
{\Large {\bf On the Corrections to Dashen's Theorem}}\\[15mm]
Robert Baur\footnote{e-mail:
\_baur@physik.unizh.ch}\\[2mm]
Institut f\"ur Theoretische Physik, Universit\"at  Z\"urich\\
CH-8057 Z\"urich, Switzerland\\[10mm]
Res Urech\footnote{e-mail:
ru@ttpux2.physik.uni-karlsruhe.de}\\[2mm]
Institut f\"ur Theoretische Teilchenphysik, Universit\"at
Karlsruhe\\
D-76128 Karlsruhe, Germany\\[18mm]
{\Large {\bf Abstract}}\\
\end{center}
The electromagnetic corrections to the masses of the pseudoscalar
mesons $\pi$ and $K$ are considered. We calculate in chiral
perturbation theory the contributions which arise from resonances
within a photon loop at order $O(e^2 m_q)$. 
Within this approach we find rather moderate deviations to Dashen's
theorem.\\[5mm]
PACS number(s): 12.39.Fe, 13.40.Dk, 14.40.Aq

\newpage
\pagestyle{plain}

\section{Introduction}

Dashen's theorem \cite{dash69} states that the squared mass
differences between the charged pseudoscalar mesons $\pi^\pm,
K^\pm$
and their corresponding neutral partners $\pi^0, K^0$ are equal in
the chiral limit, i.e., $\Delta M^2_K - \Delta M^2_\pi
= 0$, where $\Delta M^2_P = M^2_{P^\pm} - M^2_{P^0}$. In
recent years several groups have calculated the electromagnetic
corrections to this relation from non-vanishing quark masses. The
different conclusions are either  that the violation is large
\cite{dono93,bij93} or that it {\it may be} large 
\cite{mal90,ure95,neu94}.
\newline
The electromagnetic mass difference of the pions $\Delta M^2_\pi$
has
been determined in the chiral limit using current algebra by Das et
al. \cite{das67}. Ecker et al. \cite{eck89} have repeated the
calculation in the framework of chiral perturbation theory
($\chi$PT)
\cite{ga84} by resonance exchange within a photon loop. The
occurring divergences from these loops
are absorbed by introducing an electromagnetic counterterm (with a
coupling  constant $\hat{C}$) in the chiral lagrangian. They find
that
the contribution from  the loops is numerically very close to the
experimental mass difference, and thus conclude that the finite
part
of $\hat{C}$ is almost zero.\newline
In \cite{dono93} the authors have calculated the Compton scattering
of the
pseudoscalar mesons including the resonances and determined from
this amplitude the mass differences at order $O(e^2m_q)$. They concluded
first of all 
by using three low-energy relations that the one-loop result is
finite,
i.e., there is no need of a counterterm lagrangian at order $O(e^2
m_q)$ in
order to renormalize the contributions from the resonances,
secondly they
found a strong violation of Dashen's theorem. We are in
disagreement with
both of these results.\newline
In this article we proceed in an analogous manner to \cite{eck89} 
for the case
$m_q\neq 0$. We calculate in $\chi$PT the contributions of order
$O(e^2 m_q)$ to the masses of
the Goldstone bosons due to resonances. The divergences are
absorbed in the
corresponding electromagnetic counterterm lagrangian, associated
with the
couplings $\hat{K}_i$, where $i=1,\ldots,14$. The most general form
of
this
lagrangian has been given in \cite{ure95,neu94,bau95}. We find
again that
the contribution from the loops reproduces the measured mass
difference
$\Delta M^2_\pi$  very well, 
and therefore we consider the finite parts of the $\hat{K}_i$ 
to be small. Using this assumption also for the calculation of
$\Delta M^2_K$, we may
finally read off the corrections to Dashen's theorem from one-loop
resonance exchange. 
The (scale dependent) result shows that the resonances lead to rather
moderate deviations.\newline  
The article is organized as follows. In section 2 we present the
ingredients from $\chi$PT and the resonances needed for
the calculation. In section 3 we give the contributions to the
masses
and to Dashen's theorem and renormalize the counterterm lagrangian.
The
numerical results and a short conclusion are given in section 4.

\pagestyle{plain}

\section{The Lagrangians at lowest and next-to-leading Order}

The chiral lagrangian can be expanded in derivatives of the
Goldstone
fields
and in the masses of the three light quarks. The power counting
is established in the following way: The Goldstone fields are of
order
$O(p^0)$, a derivative $\partial_\mu$, the vector and axial vector
currents
$v_\mu, a_\mu$  count as quantities of $O(p)$ and the scalar
(incorporating the masses) and pseudoscalar currents $s,p$ are of
order
$O(p^2)$. The effective lagrangian starts at $O(p^2)$,
denoted by ${\cal L}_2$. It is the non-linear $\sigma$-model
lagrangian
coupled
to
external fields, respects chiral symmetry $SU(3)_{R}\times
SU(3)_{L}$, and is invariant under $P$ and $C$ transformations
\cite{ga84},
\begin{eqnarray}\label{l2}
{\cal L}_2 &=&\frac{F^2_0}{4}\langle d^\mu U^\dagger d_\mu U + \chi
U^\dagger +
\chi^\dagger U\rangle\nonumber \\
d_{\mu} U &=& \partial_{\mu} U - i(v_{\mu} + a_{\mu}) U + i U
(v_{\mu} -
a_{\mu})\nonumber \\
v_{\mu}&=&QA_{\mu} + \cdots\nonumber \\
Q&=&\frac{e}{3}\;{\rm diag}\,(2,-1,-1)\nonumber \\
\chi&=&2B_0(s+ip)\nonumber \\
s&=&{\rm diag}\,(m_u,m_d,m_s)\nonumber \\
F_\pi&=&F_0 \left[1+O(m_q )\right]\nonumber \\
B_0&=&-\frac{1}{F^2_0}\langle 0|\bar{u}u|0\rangle\left[1+
O(m_q )\right] \quad .
\end{eqnarray}
The brackets $\langle\cdots\rangle$ denote the trace in flavour
space and
$U$ is a unitary $3\times 3$ matrix that incorporates the fields of
the
eight pseudoscalar mesons,
\begin{eqnarray}
U&=&\exp\,\left(\frac{i\Phi}{F_0}\right)\nonumber \\[4mm]
\Phi&=&\sqrt{2}\left(
\begin{array}{ccc}
\frac{1}{\sqrt{2}}\pi^0+\frac{1}{\sqrt{6}}\eta_8 & \pi^+ & K^+\\
\pi^- & -\frac{1}{\sqrt{2}}\pi^0+\frac{1}{\sqrt{6}}\eta_8 & K^0\\
K^- & \overline{K^0} & -\frac{2}{\sqrt{6}}\eta_8
\end{array}
\right) \quad .
\end{eqnarray}
Note that the photon field $A_{\mu}$ is incorporated in the vector
current $v_\mu$. The corresponding kinetic term has to be added to
${\cal L}_2$,
\begin{equation}
{\cal L}^\gamma_{kin} = -\frac{1}{4} F_{\mu\nu} F^{\mu\nu} -
\frac{1}{2}
\left(
\partial_\mu A^\mu \right)^2\, ,
\end{equation}
with  $F_{\mu\nu}=\partial_\mu A_\nu -\partial_\nu A_\mu$ and the
gauge
fixing parameter chosen to be $\lambda =1$. In order to maintain
the usual
chiral counting in ${\cal L}_{kin}^\gamma$,
it is convenient to count the photon field as a
quantity of order $O(p^0)$, and the electromagnetic coupling $e$
of $O(p)$ \cite{ure95}.
\newline
The lowest order couplings of the pseudoscalar mesons to the
resonances are
linear in the resonance fields and start at order $O(p^2)$
\cite{eck89,eck289}. For the description of the fields we use the
antisymmetric tensor notation for the vector and axialvector
mesons, e.g., the vector octet has the form
\begin{equation}
V_{\mu\nu}=\left(
\begin{array}{ccc}
\frac{1}{\sqrt{2}}\rho^0_{\mu\nu}+
\frac{1}{\sqrt{6}}\omega_{8\,\mu\nu} &
\rho^+_{\mu\nu} & K^{\ast\,+}_{\mu\nu}\\[2mm]
\rho^-_{\mu\nu} &
-\frac{1}{\sqrt{2}}\rho^0_{\mu\nu}+
\frac{1}{\sqrt{6}}\omega_{8\,\mu\nu} &
K^{\ast\,0}_{\mu\nu} \\[2mm]
K^{\ast\,-}_{\mu\nu} & \overline{K^{\ast\,0}}_{\mu\nu} &
-\frac{2}{\sqrt{6}}\omega_{8\,\mu\nu}
\end{array}
\right) \quad .
\end{equation}
This method is discussed in detail in \cite{eck89}, we restrict
ourselves
on
the formulae  needed for the calculations in the following
section. The relevant interaction lagrangian contains the octet
fields
only,
\begin{eqnarray}\label{lr}
{\cal L}^V_2&=&\frac{F_V}{2\sqrt{2}}\langle
V_{\mu\nu}f_+^{\mu\nu}\rangle
+\frac{iG_V}{2\sqrt{2}}\langle
V_{\mu\nu}[u^\mu,u^\nu]\rangle\nonumber \\
{\cal L}^A_2&=&\frac{F_A}{2\sqrt{2}}\langle
A_{\mu\nu}f_-^{\mu\nu}\rangle\nonumber \\
f_\pm^{\mu\nu}&=&u F^{\mu\nu}_L u^\dagger \pm u^\dagger
F^{\mu\nu}_R u\nonumber \\
F^{\mu\nu}_{R,L}&=&\partial^\mu (v^\nu\pm a^\nu)
-\partial^\nu(v^\mu\pm
a^\mu) - i[v^\mu\pm a^\mu,v^\nu\pm a^\nu]\nonumber \\
u^\mu&=&iu^\dagger d^\mu U u^\dagger =u^{\dagger\,\mu}\nonumber \\
U&=&u^2 \quad .
\end{eqnarray}
The coupling constants are real and are not restricted by chiral
symmetry
\cite{eck289}, numerical estimates are given in \cite{eck89}. In
the kinetic
lagrangian a covariant derivative acts on the vector and
axialvector mesons,
\begin{eqnarray}\label{rkin}
{\cal L}^R_{kin}&=&-\frac{1}{2}\langle \nabla^\mu R_{\mu\nu}
\nabla_\sigma R^{\sigma\nu} - \frac{1}{2}M^2_R
R_{\mu\nu}R^{\mu\nu}\rangle\hspace{2cm}R=V,A\nonumber \\
\nabla^\mu R_{\mu\nu}&=&\partial^\mu R_{\mu\nu} +
[\Gamma^\mu,R_{\mu\nu}]\nonumber \\
\Gamma^\mu&=& \frac{1}{2}\left\{u^\dagger [\partial^\mu
-i(v^\mu + a^\mu)]u + u[\partial^\mu -i(v^\mu - a^ \mu)]u^\dagger
\right\}\quad,
\end{eqnarray}
where $M_R$ is the corresponding mass in the chiral limit. Finally
we
collect all the different terms together into one lagrangian,
\begin{equation}\label{resonance}
{\cal L}^{eff}_2={\cal L}_2 + {\cal L}^R_2 + {\cal L}^\gamma_{kin}
+ {\cal
L}^R_{kin} \quad
{}.
\end{equation}
The one-loop electromagnetic mass shifts of the pseudoscalar mesons
calculated with this lagrangian (see section 3) contain divergences
that
can be absorbed in a counterterm lagrangian. In its general form,
this
lagrangian has one term of order $O(e^2)$ and 14 terms of
$O(e^2 p^2)$ \cite{ure95,neu94,bau95},
\begin{eqnarray}\label{l4}
{\cal L}^C_2&=&\hat{C}\langle Q U Q U^\dagger\rangle\nonumber \\
{\cal L}^C_4&=&\hat{K}_{1}F^2_0\langle d^\mu U^\dagger d_\mu
U\rangle
\langle Q^2\rangle
+\hat{K}_{2}F^2_0 \langle d^\mu U^\dagger d_\mu U\rangle
\langle Q U Q U^\dagger\rangle \nonumber \\
&&+\hat{K}_{3} F^2_0\left(\langle d^\mu U^\dagger Q U\rangle
\langle d_\mu U^\dagger Q U\rangle
+\langle d^\mu U Q U^\dagger\rangle \langle d_\mu U Q
U^\dagger\rangle
\right)\nonumber \\
&&+\hat{K}_{4} F^2_0 \langle d^\mu U^\dagger Q U\rangle
\langle d_\mu U Q U^\dagger \rangle \nonumber \\
&&+\hat{K}_{5} F^2_0 \langle  \left( d^\mu U^\dagger d_\mu U
+ d^\mu U d_\mu U^\dagger\right) Q^2 \rangle\nonumber \\
&&+\hat{K}_{6} F^2_0 \langle d^\mu U^\dagger d_\mu U  Q U^\dagger Q
U +
d^\mu U
d_\mu U^\dagger Q U Q U^\dagger\rangle\nonumber \\
&&+\hat{K}_{7} F^2_0\langle \chi U^\dagger + \chi^\dagger U\rangle
\langle Q^2\rangle
+\hat{K}_{8} F^2_0 \langle \chi U^\dagger + \chi^\dagger U \rangle
\langle Q U Q U^\dagger\rangle \nonumber \\
&&+\hat{K}_{9} F^2_0 \langle (\chi U^\dagger + \chi^\dagger U
+ U^\dagger \chi + U \chi^\dagger) Q^2\rangle\\
&&+\hat{K}_{10} F^2_0 \langle (\chi^\dagger U + U^\dagger \chi) Q
U^\dagger Q U
+ (\chi U^\dagger + U \chi^\dagger ) Q U Q
U^\dagger\rangle\nonumber \\
&&+\hat{K}_{11} F^2_0 \langle (\chi^\dagger U - U^\dagger \chi)Q
U^\dagger Q U
+ (\chi U^\dagger  - U \chi^\dagger) Q U Q
U^\dagger\rangle\nonumber \\
&&+\hat{K}_{12} F^2_0 \langle  d^\mu U^\dagger \left[ c^R_\mu Q , Q
\right] U
+ d^\mu U \left[ c^L_\mu Q , Q \right] U^\dagger\rangle\nonumber \\
&&+\hat{K}_{13} F^2_0 \langle c^{R\,\mu}Q U c^L_\mu Q
U^\dagger\rangle
+\hat{K}_{14} F^2_0 \langle c^{R\,\mu} Q c^R_\mu Q + c^{L\,\mu}
c^L_\mu
Q
\rangle\nonumber \\
&&+O(p^4,e^4)\nonumber \quad .
\end{eqnarray}
with $c^{R,L}_\mu Q = -i\left[v_\mu \pm a_\mu,Q\right]$.
The three last terms contribute only to matrix elements with
external
fields, we are therefore left with 12 relevant counterterms. Note
that we
have omitted terms which come either from the purely strong or
the purely electromagnetic sector in ${\cal L}^C_4$.\newline
At this point it is
 worthwhile to discuss the connection of the present formalism to the usual
$\chi$PT without resonances where the Goldstone bosons and the (virtual)
photons are the only interacting particles. For this purpose we consider
the electromagnetic mass of the charged pion. In $\chi$PT the lagrangian
has the form up to and including $O(e^2p^2)$
\begin{eqnarray}
{\cal L} &=& {\cal L}_2^Q + {\cal L}_4^Q \nonumber \\
{\cal L}_2^Q &=& {\cal L}_2 + C\langle Q U Q U^\dagger\rangle,
\hspace{2cm}{\cal L}_4^Q = \sum^{14}_{i=1} K_i O_i
\end{eqnarray}
where $C$ and $K_i$ are low energy constants. They are independent of the
Goldstone bosons masses 
and parameterize all the underlying physics (including
resonances) of
$\chi$PT. ${\cal L}_2$ is given in (\ref{l2}) and the operators $O_i$ are
identical to those in (\ref{l4}). Neglecting the
contributions of the order $O(e^2m_q)$ for a moment, the pion mass is
\cite{eck89} 
\begin{equation}
M_{\pi^\pm}^2 = \frac{2e^2}{F_0^2}C + O(e^2m_q)
\end{equation}
entirely determined by the coupling constant $C$. 
In the resonance approach $M_{\pi^\pm}^2$ gets contributions from
resonance-photon loops already at order $O(e^2)$  (see graphs (c)
and (d) in Figure 1)
\begin{equation}
M_{\pi^\pm}^2 = M_{\pi^\pm}^2|_{loops} + \frac{2e^2}{F_0^2}\hat C 
                 + O(e^2m_q)
\end{equation}
The loop term contains a divergent and a finite part and is completely
determined by the resonance parameters. 
The divergences are absorbed by renormalizing
the coupling constants $\hat C$ \cite{eck89} (see section 3).  The
connection to $\chi$PT without resonances is then given by the relation 
\cite{eck89}
\begin{eqnarray}
C &=& C^R(\mu) + \hat C(\mu) \nonumber \\
C^R(\mu) &=& \frac{F_0^2}{2e^2}M_{\pi^\pm}^2|_{loops\;(finite)} 
\label{finite}
\end{eqnarray}
where $C^R(\mu)$ and $\hat C(\mu)$ are finite and the scale dependence
cancels in the sum. 
Relation (\ref{finite}) says that
the coupling constant $C$ is split in a part from resonances 
$(C^R)$ and another part from non-resonant physics $(\hat C)$. 
This ansatz of separating resonant and non-resonant contributions to the
low-energy parameters has been originally made for the strong interaction
sector at next to leading order \cite{eck89}. In this case
resonance 
exchange gives tree-level contributions and no renormalization is
needed. In the electromagnetic case however,
contributions arise from resonances with photons in loops and we 
renormalize the non-resonant part of the coupling constant, i.e. $\hat
C$ at order $O(e^2)$.\newline 
In an analogous fashion the above procedure can be carried out up to
the order $O(e^2 p^2)$. The couplings $K_i$ of ${\cal L}_4^Q$ are in
general divergent, since they absorb the divergences of the one-loop
functional generated by ${\cal L}_2^Q$ \cite{ure95,neu94,bau95}.
At a specific scale point the renormalized coupling constants $K_i^r(\mu)$
can  be split in two parts
\begin{equation}
K_i^r(\mu_0) = K_i^R(\mu_0) + \hat K_i(\mu_0) 
\end{equation}
where the terms on the right-hand side are taken after renormalization 
of $\hat K_i$ (see section 3) and are thus finite. 
The choice of the scale point $\mu_0$ is not a priori fixed.
Like in the strong sector \cite{eck89} we consider $\mu_0$ in the range
of the lowest lying resonances, i.e. in the range from $0.5$ to $1.0 
\mbox{ GeV}$.

In the strong sector it was found that 
the resonances saturate the low-energy parameters
almost completely \cite{eck89}. In additon the authors have found 
that the
same conclusion holds for the electromagnetic coupling constant
$C$ leading to $\hat C(\mu) \approx 0$. 
Consequently we $assume$ that the $K_i^r(\mu)$ are also
saturated by resonance contributions, i.e. we put
\begin{equation}
\hat K_i(\mu) \approx 0 \quad.     
\end{equation}
As we will see in section 4 this assumption works well in the case of
$\Delta M^2_\pi$. 

\section{Corrections to Dashen's Theorem}

Using the lagrangian given in (\ref{resonance}) it is a
straightforward
process to
calculate the mass shift between the charged pseudoscalar mesons
$\pi^\pm,
K^\pm$ and their corresponding neutral partner $\pi^0, K^0$ at the
one-loop
level.
\begin{figure}[t]
 \begin{center}
 \begin{tabular}{c}
   \epsfxsize=13.0cm
   \leavevmode
   \epsffile[67 358 546 746]{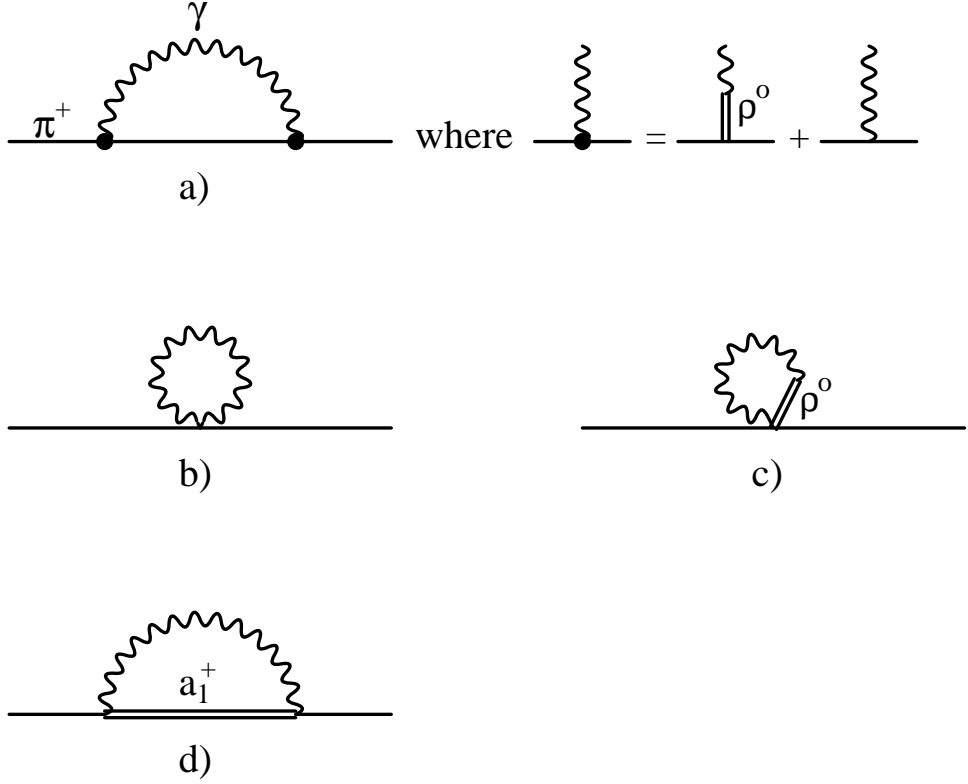}
 \end{tabular}
 \caption[]{\label{fig1}One-loop contributions to the
electromagnetic mass
shift of $\pi^\pm$.}
 \end{center}
\end{figure}
The relevant diagrams for the mass of the charged pion are shown in
Fig.1. Graph (a) contains
the off-shell pion form factor, (b) vanishes in
dimensional regularization and  (c) is called ``modified seagull
graph''. Graph (d) contains an $a_1$-pole. The mass of the neutral
pion
does not get contributions  from the loops.\newline
If we take the resonances to be in the $SU(3)$ limit according to
(\ref{rkin}), i.e., all vector resonances
have the same mass $M_V $ and all axialvector resonances the mass
$M_A$, we get the contributions listed below. For the graphs
with the pion form factor,
\begin{eqnarray}
\Delta_{p.f.}M^2_\pi &=& -ie^2 \int \frac{d^4 q}{(2\pi)^4}
\frac{q^2+4\nu+4M^2_\pi}{q^2(q^2+2\nu)}\nonumber
\\
&& -i\frac{8e^2 F_V G_V}{F^2_0} \int \frac{d^4q}{(2\pi)^4}
      \frac{q^2M^2_\pi-\nu^2}{q^2 (q^2+2\nu) (M^2_V - q^2)}\\
&& -i\frac{4e^2 F^2_V G^2_V}{F^4_0}\int \frac{d^4 q}{(2\pi)^4}
      \frac{q^2[q^2 M^2_\pi-\nu^2]}{q^2 (q^2+2\nu) (M^2_V - q^2)^2}
\nonumber \quad ,
\end{eqnarray}
where $\nu = pq$ and $p$ is the momentum of the pion. Using the
relation
$F_V G_V = F^2_0$
\cite{eck289} we obtain
\begin{equation}
\Delta_{p.f.}M^2_\pi = -ie^2 M^4_V \int \frac{d^4 q}{(2\pi)^4}
                \frac{2\nu + 4M^2_\pi }{q^2(q^2+2\nu ) (M^2_V -
q^2)^2}
\quad .
\end{equation}
The modified seagull graph gives
\begin{equation}
\Delta_{s.g.}M^2_\pi = i \frac{e^2 F^2_V }{F^2_0}  (3-\epsilon)
                      \int \frac{d^4 q}{(2\pi)^4}\frac{1}{M^2_V -
q^2}
\end{equation}
with $\epsilon = 4-d$, and finally for the $a_1$-pole graph, where
unlike
\cite{dono93} we get an additional second term,
\begin{eqnarray}\label{forgot}
\Delta_{a_1}M^2_\pi &=& -i \frac{e^2 F^2_A}{F^2_0}(3-\epsilon)
                         \int \frac{d^4
q}{(2\pi)^4}\frac{1}{M^2_A-q^2}
\nonumber \\
&-& i \frac{e^2 F^2_A}{F^2_0}
        \int \frac{d^4 q}{(2\pi)^4}\frac{q^2\left[M^2_\pi
                  +(3-\epsilon)\nu \right]
        +(2-\epsilon)\nu^2}{q^2\left[M^2_A-(q+p)^2\right]} \quad .
\end{eqnarray}
We now add the contribution from ${\cal L}^C_2$ and ${\cal L}^C_4$ 
to the mass shift 
\cite{ure95,neu94} and evaluate the integrals,
\begin{eqnarray}\label{div}
\Delta M^2_\pi &=& -\frac{3e^2}{F^2_0 16\pi^2}
      \left[ F^2_V M^2_V \left(\ln\frac{M^2_V}{\mu^2}
               + \frac{2}{3}\right)
            -F^2_A M^2_A\left(\ln\frac{M^2_A}{\mu^2}
               + \frac{2}{3}\right)\right]\nonumber \\
&& -\frac{e^2 F^2_A}{F^2_0 16\pi^2} M^2_\pi
    \left[2+\frac{3}{2}\ln\frac{M^2_A}{\mu^2}
    +I_1\left(\frac{M^2_\pi}{M^2_A}\right)\right]\nonumber \\
&& +\frac{2e^2}{16\pi^2}M^2_\pi
         \left[\frac{7}{2}-\frac{3}{2}\ln\frac{M^2_\pi}{M^2_V }
         +I_2\left(\frac{M^2_\pi}{M^2_V}\right)\right] \\
&& +\frac{2e^2\hat{C}}{F^2_0} - \frac{6e^2}{F^2_0}
                       (F^2_V M^2_V - F^2_A M^2_A)\lambda\nonumber
\\
&& +8e^2M^2_K \hat{K}_{8}+2e^2M^2_\pi \hat{R}_{\pi}
 	 -\frac{3e^2 F^2_A}{F^2_{0}}M^2_\pi\lambda\nonumber
\end{eqnarray}
with
\begin{eqnarray}
I_1(z) &=& \int^1_0 x\ln[x-x(1-x)z] \,dx\nonumber\\
I_2(z) &=&  \int^1_0 (1+x)\left\{\ln[x+(1-x)^2 z]
                   -\frac{x}{x+(1-x)^2z}\right\}\,dx\nonumber \\
\hat{R}_\pi &=& -2 \hat{K}_3 + \hat{K}_4 + 2\hat{K}_8 +
4\hat{K}_{10} + 4\hat{K}_{11}
\end{eqnarray}
The divergences of the resonance-photon loops show up as poles
 in $d=4$ dimensions. They are collected in the terms
proportional to $\lambda$
\begin{eqnarray}
\lambda &=& \frac{\mu^{d-4}}{16\pi^2} \left\{\frac{1}{d-4}
                   -\frac{1}{2}[\ln 4\pi+\Gamma'(1) +1 ]
\right\}\quad .
\end{eqnarray}
The occurring divergences  are now canceled
by renormalizing the contributions from non-resonant physics, i.e. the
coupling constants $\hat C$ and $\hat K_{i}$. The divergence of the order
$O(e^2)$ (fourth
line of equation (\ref{div})) is absorbed by putting
\begin{eqnarray}
\hat C &=& \hat C(\mu)+3(F^{2}_{V}M^{2}_V-F^{2}_{A}M^{2}_{A})\lambda 
\label{div1}
\end{eqnarray}
and that of the order $O(e^2 m_{q})$ (fifth line of equation 
(\ref{div})) by the relation
\begin{eqnarray}
\hat R_{\pi} &=& \hat R_{\pi}(\mu)+\frac{3F^{2}_{A}}{2F^2_0}\lambda \quad.
\label{div2}
\end{eqnarray}
Using the second Weinberg sum rule \cite{wein67}
\begin{equation}\label{wein2}
F^2_V M^2_V - F^2_A M^2_A=0\quad ,
\end{equation}
the divergence in (\ref{div1}) cancels, but the divergence in
(\ref{div2}) does not. Even if we used an extension of this sum
rule to
order $O(m_q)$ \cite{pas82},
\begin{equation}
F^2_\rho M^2_\rho -F^2_{a_1}M^2_{a_1}\simeq F^2_\pi M^2_\pi\quad
\end{equation}
and assumed $F_A = F_0$ \cite{eck289}, the divergence would not
cancel,
on the contrary, it would become larger.\newline
We finally get the result
\begin{eqnarray} \label{finalpi}
\Delta M^2_\pi &=&-\frac{3e^2}{F^2_0 16\pi^2}
F^2_V M^2_V \ln\frac{M^2_V}{M^2_A}\nonumber\\
&& -\frac{e^2F^2_A}{F^2_0 16\pi^2} M^2_\pi
    \left[2+\frac{3}{2}\ln\frac{M^2_A}{\mu^2}
     +I_1\left(\frac{M^2_\pi}{M^2_A}\right)\right]\nonumber \\
&& +\frac{2e^2}{16\pi^2}M^2_\pi
    \left[\frac{7}{2}-\frac{3}{2}\ln\frac{M^2_\pi }{M^2_V}
     +I_2\left(\frac{M^2_\pi}{M^2_V}\right)\right]\\
&& +\frac{2e^2\hat{C}}{F^2_0}
    +  8e^2 M^2_K \hat{K}_8+2e^2 M^2_\pi \hat{R}_\pi
(\mu)\nonumber\quad ,
\end{eqnarray}
where we used (\ref{wein2}) to simplify the first term. In the
chiral limit
$\Delta M^2_\pi$ reduces to the expression given in
\cite{eck89}.\newline
The mass difference for the kaons is determined in an analogous
way, in the
contribution from the loops we merely have to replace $M^2_\pi$ by
$M^2_K$.
Finally the formula for the corrections to Dashen's theorem may be
read
off,
\begin{eqnarray}\label{dashen}
\Delta M^2_K - \Delta M^2_\pi &=&
- \frac{e^2 F^2_A}{F^2_0 16\pi^2}
  \left\{M^2_K\left[2+\frac{3}{2}\ln\frac{M^2_A}{\mu^2}
        +I_1\left(\frac{M^2_K}{M^2_A}\right)\right]\right.\nonumber
\\
&&\hspace{2.2cm}\left.- M^2_\pi\left[2
             +\frac{3}{2}\ln\frac{M^2_A}{\mu^2}
+I_1\left(\frac{M^2_\pi}{M^2_A}\right)\right]\right\}\nonumber \\
&&+ \frac{2e^2}{16\pi^2}\left\{ M^2_K
    \left[\frac{7}{2}-\frac{3}{2}\ln\frac{M^2_K}{M^2_V}
     +I_2\left(\frac{M^2_K}{M^2_V}\right)\right]\right.\nonumber \\
&&\hspace{1.7cm}\left. - M^2_\pi
    \left[\frac{7}{2}-\frac{3}{2}\ln\frac{M^2_\pi}{M^2_V}
     +I_2\left(\frac{M^2_\pi}{M^2_V}\right)\right]\right\}\\
&&- 2e^2 M^2_K \left[\frac{2}{3}\hat{S}_K(\mu) + 4\hat{K}_8\right]
    + 2e^2 M^2_\pi\left[\frac{2}{3}\hat{S}_\pi -
\hat{R}_\pi(\mu)\right]\quad
,\nonumber
\end{eqnarray}
where $\hat{S}_{\pi,K}$ represent the contributions from the
counterterm
lagrangian to $\Delta M^2_K$,
\begin{eqnarray}
\hat{S}_\pi &=& 3\hat{K}_8 + \hat{K}_9 + \hat{K}_{10}\nonumber \\
\hat{S}_K &=& \hat{K}_5 + \hat{K}_6 - 6\hat{K}_8 - 6\hat{K}_{10} -
6\hat{K}_{11}\nonumber \\
\hat{S}_K &=& \hat{S}_K(\mu) + \frac{3F^2_A}{2F^2_0}\lambda
\quad .
\end{eqnarray}

\section{Numerical Results and Conclusion}

We put $F_0$ equal to the physical pion decay constant, $F_\pi =
92.4 \;{\rm MeV}$
and the masses of the mesons to $M_\pi = 135 \;{\rm MeV}, M_K = 495
\;{\rm MeV}$.
We take
$F_V = 154 \;{\rm MeV}$ \cite{eck89} and $M_V = M_\rho = 770 \;{\rm
MeV}$. To
eliminate the parameters of the axialvector resonances we use
Weinberg's
sum rules \cite{wein67},
\begin{equation}
F^2_V - F^2_A = F^2_0\hspace{2cm}
F^2_V M^2_V - F^2_A M^2_A=0 \quad .
\end{equation}
The contributions from the counterterm lagrangian are not known so far. 
In \cite{eck89} it was found that the experimental mass difference 
$\Delta M^2_\pi$ at order $O(e^2)$ is well reproduced by the
resonance-photon loops and therefore the authors conclude that the
contributions from non-resonant physics are small, i.e. 
$\hat{C}\approx 0$. In analogy we $assume$ for the numerical 
evaluation the
dominance of the resonant contributions at order $O(e^2m_q)$, i.e. 
we put $\hat{K}_i(\mu)\approx 0$.\newline
Putting the numbers in
(\ref{finalpi}) we get for the contribution from the loops to
$\Delta M^2_\pi$
at the scale points $\mu=(0.5,0.77,1) \;{\rm GeV}$ (see Fig.2a)
\begin{equation}\label{numpi}
\Delta M^2_\pi |_{loops}= 2M_\pi\times (\;5.0\, ,\;5.1\, ,\;5.1\;)
\;{\rm MeV}\quad .
\end{equation}
which is in nice agreement with the experimental value $\Delta M^2_\pi
|_{exp.} = 2M_\pi\times 4.6\;{\rm MeV}$ \cite{pdg94}. 
Using resonance saturation  in the Kaon system as well,
we obtain for the corrections to Dashen's
theorem (again at
the scale points $\mu=(0.5,0.77,1) \;{\rm GeV}$)
\begin{equation}
\Delta M^2_K - \Delta M^2_\pi = (\;-0.13\, ,\;0.17\,
,\;0.36\;)\times10^{-3}\;({\rm GeV})^2
\quad ,
\end{equation}
which are smaller than the values found in the literature,
\begin{eqnarray}
\Delta M^2_K - \Delta M^2_\pi=\left\{
\begin{array}{ccc}
1.23&&\cite{dono93}\\
1.3\pm 0.4&\times 10^{-3} \;({\rm GeV})^2&\cite{bij93}\\
0.55\pm 0.25&&\cite{mal90}\\
\end{array}
\right.
\end{eqnarray}
\begin{figure}[t]
 \begin{center}
 \begin{tabular}{lr}
   \epsfxsize=6.6cm
   \leavevmode
   \epsffile[18 144 593 718]{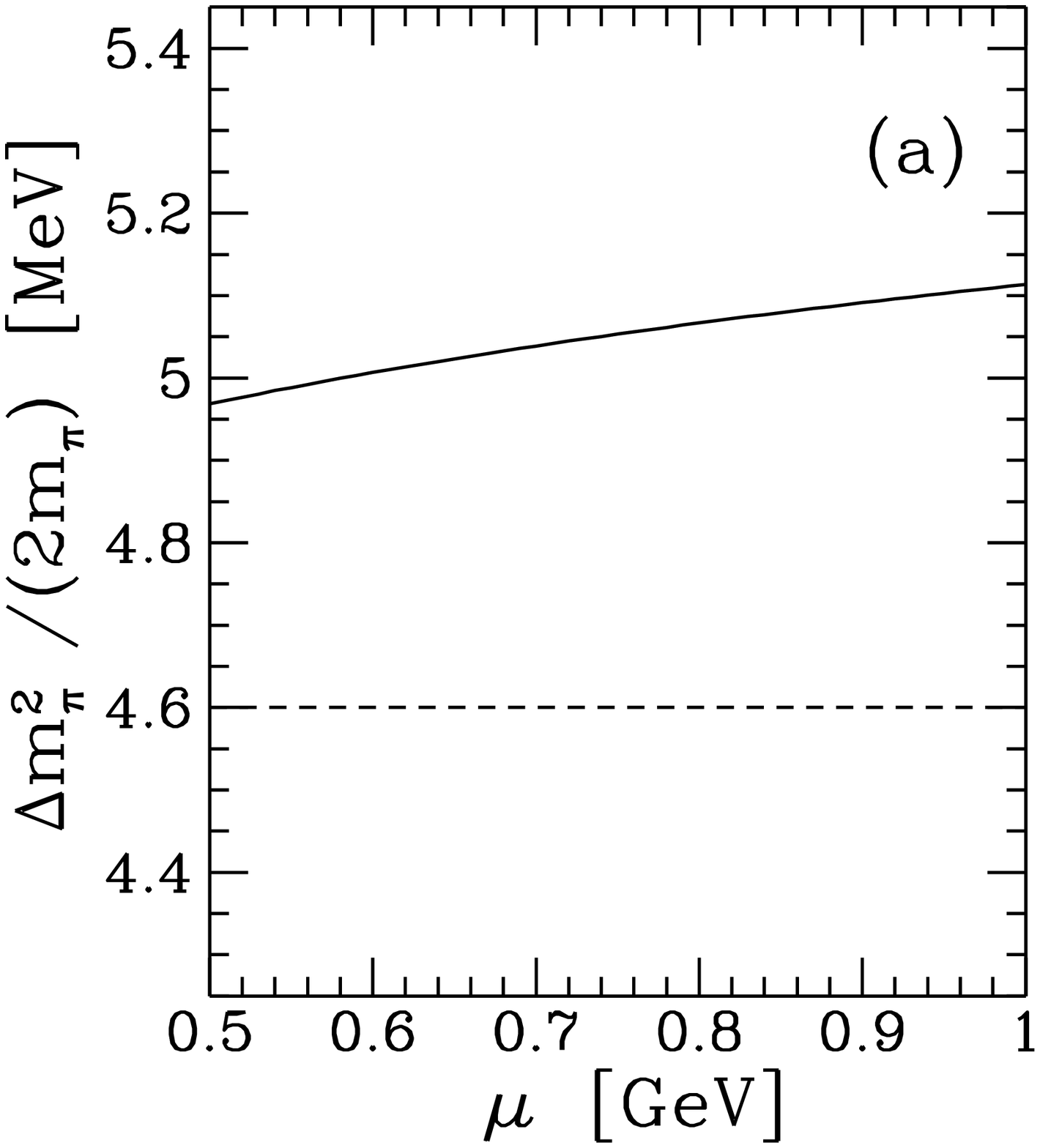} &
   \epsfxsize=6.6cm
   \leavevmode
   \epsffile[18 144 593 718]{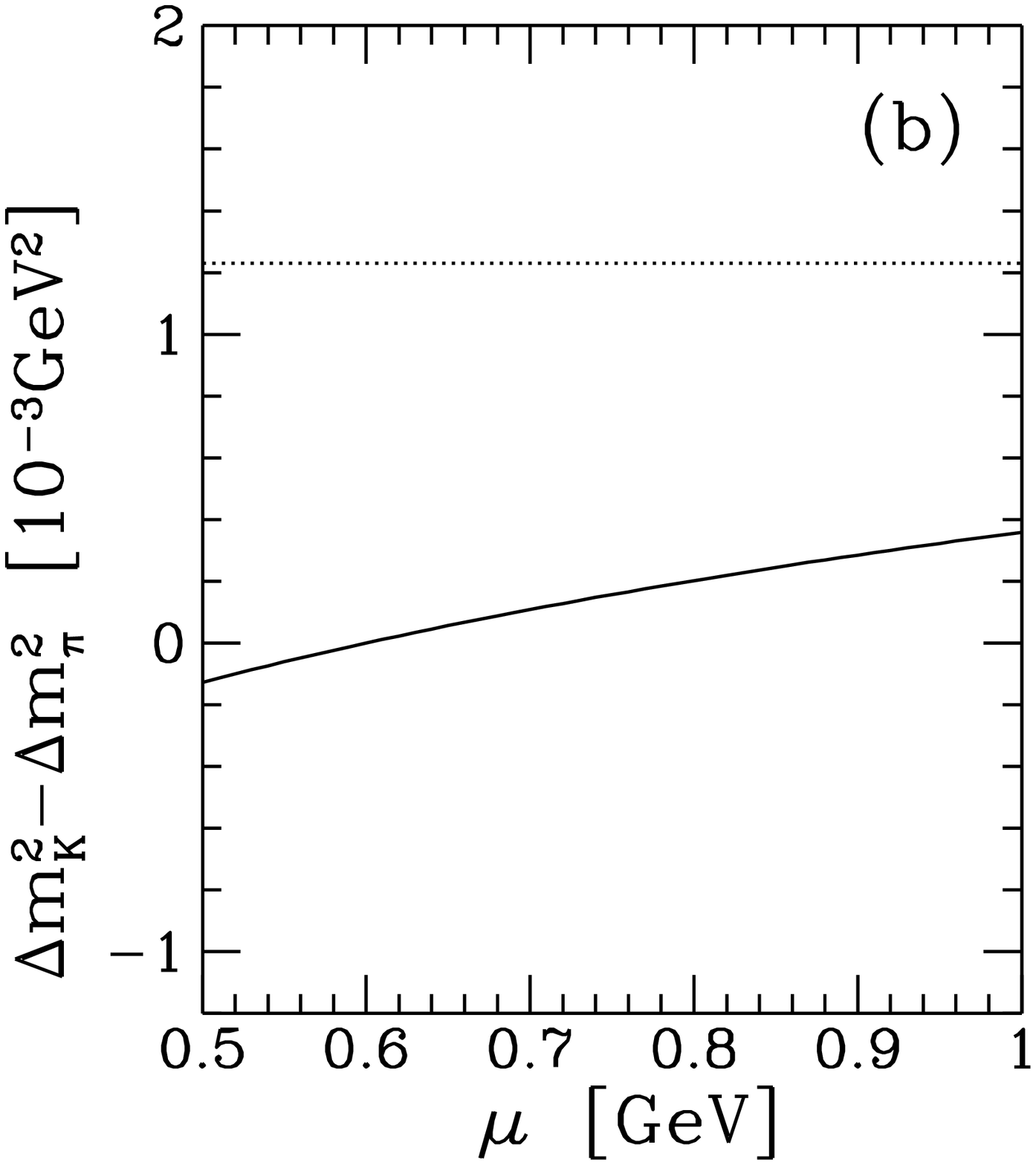}
 \end{tabular}
 \caption[]{\label{fig2} The solid lines show our results, the
dashed and
dotted curves represent in a) the experimental value \cite{pdg94},
in b)
the result of \cite{dono93}, respectively.}
 \end{center}
\end{figure}
Of course, in order to get a scale independent result, the
counterterms are not allowed to vanish completely. 

In \cite{dono93} the authors  calculated the Compton scattering
of the
Goldstone bosons within the same model that
we have used in the present article and  determined the corrections
to Dashen's theorem by closing the photon line. Their calculation
is finite
(without counterterms) and gives a considerably large value for
$\Delta M^2_K
- \Delta M^2_\pi$. The difference to our result may be
identified in (\ref{forgot}), where we have found an additional
(singular) term
that gives a large negative and scale dependent contribution. The
two
results are compared in
Fig.2b. Note that in \cite{dono93} the physical masses for
the
resonances
are used  in the calculation of $\Delta M^2_K$, whereas we
work in the $SU(3)$ limit throughout.\newline
The other calculations are not strongly connected to our approach,
for a discussion of the value given in \cite{mal90} we refer to
\cite{ure95}.\newline
We therefore conclude that taking into account the resonances
at the one-loop level and working strictly in the $SU(3)$ limit for the
resonances leads to moderate rather than large corrections to Dashen's
theorem. Possibly strong violations must come from higher loop
corrections or from non-resonant  physics.\\[8mm]
{\Large {\bf Acknowledgements}}\\[2mm]
We thank G.Ecker, J.Gasser, J.Kambor, H.Leutwyler and D.Wyler for
helpful
discussions.

\end{document}